\documentclass[preprint,square,12pt]{elsarticle}
\usepackage{amssymb,dcolumn,bm,mathrsfs,epsfig,psfrag,color,times}
\journal{Physics Letters A}
\begin{document}
\begin{frontmatter}
\title{Exact Solutions of a Spin-Orbit Coupling Model in Two-Dimensional
Central-Potentials and Quantum-Classical Correspondence}
\author[label1,label2]{Jun-Li Xin}
\author[label1]{J.-Q. Liang\corref{cor1}}
\cortext[cor1]{Corresponding author:J.-Q. Liang Tel.: +86 351 7011399; fax: +86
351 7011399.} \ead {jqliang@sxu.edu.cn}
\address[label1]{Institute of Theoretical Physics, Shanxi University, Taiyuan, Shanxi 030006, China}
\address[label2]{Department of Physics and Electronic Engineering, Yuncheng University,Yuncheng, 044000,China}
\date{\today}
\begin{abstract}
In this paper we present both the classical and quantum periodic-orbits of a
neutral spinning particle constrained in two-dimensional central-potentials
with a cylindrically symmetric electric-field in addition which leads to an
effective non-Abelian gauge field generated by the spin-orbit coupling.
Coherent superposition of orbital angular-eigenfunctions obtained explicitly
at the condition of zero-energy exhibits the quantum-classical correspondence
in the meaning of exact coincidence between classical orbits and spatial
patterns of quantum wave-functions, which as a consequence results in the
fractional quantization of orbital angular-momentum by the requirement of the
same rotational symmetry of quantum and classical orbits. A non-Abelian
anyon-model emerges in a natural way.

\end{abstract}

\begin{keyword}
Spin-Orbit coupling \sep Non-Abelian gauge field \sep Quantum-Classical Correspondence \sep Fractional quantization of orbital angular-momentum\\
\PACS 03.65.Ge \sep 05.30.Pr \sep 42.60.Jf \sep 03.65.Vf
\end{keyword}

\end{frontmatter}

\section{Introduction}

Quantum-classical correspondence (QCC) in one-dimensional space is a well
known issue, for example, the quantum dynamics of harmonic oscillator in
coherent states coincides exactly with the classical one. While it has
attracted considerable interesting just recently for two-dimensional (2D)
harmonic oscillator\cite{Chen}, central potentials \cite%
{Nieto,Makowski2,Makowski3} and for the non-Hermitian many-particle system,
namely, a non-Hermitian N-particle Bose-Hubbard dimer with a complex on-site
energy \cite{Graefe}. The QCC in stationary states, which means quite
naturally and inevitably that quantum wave functions are localized on
classical orbits, plays a central role in the explanation of many peculiar
quantum-phenomena, for example, shell effects in nuclei and metallic
clusters \cite{Brack,De}, fluctuations of conductance in mesoscopic
semiconductor billiards \cite{Zozoulenko,Brunner}, and the oscillations of
photodetachment cross-sections \cite{Peters,Bracher} as well.

In the interesting works\cite{Nieto,Makowski2,Makowski3,Makowski1} both the
classical and quantum periodic-orbits of a particle in 2D
scalar-central-potentials of the general form are obtained analytically with
the zero total-energy and it is shown that the fractional
angular-momentum-quantization can be determined by the QCC with the wave
functions well localized on the classical periodic-orbits, which imposes a
special boundary condition on the angular wave functions \cite%
{Makowski2,Makowski3,Makowski1}. The zero-energy states are of importance
for the description of cold-atom collisions \cite{Sadeghpour,Wang} and some
vortex lattices \cite{Kobayashi}. The spin coherent states \cite%
{E,Zhang,J.R,Kh} which are the most representative states related to
classical dynamics and the most persistent states describing the interaction
with the environment\cite{Zurek} are essential in the construction of
quantum wave functions localized on the classical orbits in 2D
central-potentials and the 2D harmonic oscillators\cite{Chen} as well.

The spin-orbit (SO) interaction, which coupling the internal and orbital
degrees of freedom exists intrinsically in semiconductors, is responsible
for the quantum spin-Hall-effect\cite{Mura,Kato} and has been realized
experimentally for pseudo spin-1/2 atomic-condensates recently \cite%
{Ho,Lin,Yip}. It has become a very active field of research known as
spintronics\cite{Wolf}, where the spin degree of freedom of the electron is
exploited for improved functionality of electronic devices. Moreover, the SO
is crucial in the description of tunneling magnetoresistance\textbf{\ }\cite%
{Matos} and also of cold atom-dynamics\cite{Jacob}. SO coupling is naturally
depicted as the interaction between an effective non-Abelian gauge potential
and a particle with spin. In quantum systems, the concept of non-Abelian
gauge-field was proposed by Wilczek and Zee in the context of Berry phase
for a system of degenerate ground-states \cite{Wilczek}. The realization of
non-Abelian gauge fields in cold atomic systems was theoretically suggested
for a neutral spinning particle (pseudo-spin) in electromagnetic fields \cite%
{Dum} as well as in a cavity QED system \cite{Larson}.

As of yet the QCC in the meaning that the density of wave functions is well
localized on the classical orbits has been studied only for the point
particle without spin, since the spin is directly introduced in
non-relativistic quantum mechanics and has no classical counterpart. The
dynamics for a particle with intrinsic angular momentum or spin is different
from that of a point particle and the key point is the definition of
classical spin-variable \cite{Bala}, based on which the both classical and
quantum dynamics of a neutral spinning-particle in the Aharonov-Casher (AC)
gauge field\cite{Aharonov} was explored long ago \cite{Liang2} giving rise
to the theoretical analyses of neutron-experiments for the Aharonov-Bohm
effect \cite{Liang1}. We in the present paper study the QCC for a spinning
particle confined in 2D central-potentials, in which macroscopic wave
functions are constructed by the coherent superposition of orbital
angular-eigenfunctions in the spirit of usual SU(2) spin coherent-states
\cite{E,Zhang,J.R,Kh}. Both classical orbits and quantum probability-clouds
in the macroscopic states are obtained, which possess the exactly same
spatial patterns. Moreover, it is shown that the SO coupling leads naturally
to a non-Abelian gauge potential and acts as a non-Abelian anyon model,
which may be realized with cold-atoms in planner traps or with planner
quantum-dots.

The classical dynamics of spinning particle with spin treated as a classical
variables is formulated and general solutions of periodic-orbits are
obtained in Sec. II. In Sec. III, the QCC is established by the coherent
superposition of orbital angular-eigenfunctions, which along with the
radial-part solutions of the Schr$\ddot{o}$dinger equation form the desired
probability clouds of wave functions well localized on classical periodic
orbits. In addition both the classical and quantum precessions of spin
variables are also demonstrated explicitly.\emph{\ }Our solutions give rise
to the explicit non-Abelian gauge potential and the property of non-Abelian
anyon.

\section{Classical dynamics of spin and periodic orbits}

\subsection{Lagrangian and Hamiltonian}

\textbf{\ \ }The classical dynamics of a neutral spinning-particle
constrained in the 2D central-potential of $V(r)=-\frac{\varrho }{r^{2k+2}}$%
, with ${r}=\sqrt{x^{2}+y^{2}}$ and $\varrho $ being a potential constant,
can be described by the Lagrangian
\begin{equation}
\mathcal{L}%
 _{0}=\frac{1}{2}[M\mathbf{\dot{r}}^{2}+i\lambda Tr(\sigma
_{3}g^{-1}\dot{g})]-V(r)
\end{equation}%
where $M$ and $\lambda $ are mass and spin-value of the particle. The
configuration space is a product space\textbf{\ }$%
\mathbb{R}
^{2}\times \Gamma $\textbf{, }where\textbf{\ }$\Gamma =\{g\}$\textbf{\ }%
denotes the usual\textbf{\ }spin\textbf{-}$\frac{1}{2}$\textbf{\ }%
representation of the rotation group\textbf{\ }with element $g$\textbf{\ }%
being a $SU(2)$ matrix \cite{Bala,Liang1},\textbf{\ }$g^{\dagger }g=1,\det
g=1.$ The (classic)\textbf{\ }dynamic variables of spin\textbf{\ }$\mathbf{S}
$ constrained by\textbf{\ }$\mathbf{S}^{2}=S_{1}^{2}+S_{2}^{2}+S_{3}^{2}=%
\lambda ^{2}$ are connected with the group element $g$\ through the relation%
\textbf{\ }%
\begin{equation}
\mathbf{S\cdot \mbox{\boldmath$\sigma$}}=\lambda g\sigma _{3}g^{-1}
\end{equation}%
where $\sigma _{i}$ is\textbf{\ }Pauli matrices\textbf{.} We moreover assume
that a cylindrically symmetric electric field
\begin{equation}
\mathbf{E=}\frac{\eta \mathbf{e}_{\mathbf{r}}}{r}  \label{3}
\end{equation}%
is applied, which can be realized by an infinitely long line-charge with
charge-density (the charge of per unit length) $\eta $. The interaction part
of Lagrangian between the moving spin of velocity $\mathbf{\dot{r}}$ and the
electric field, i.e. the SO, is \textbf{\ }%
\begin{equation}
\mathcal{L}%
_{i}=\frac{\mu \lambda }{2c}Tr[g\sigma _{3}g^{-1}(\mathbf{%
E\times \mathbf{\dot{r}}})\cdot \mathbf{\mbox{\boldmath$\sigma$}}]
\end{equation}%
where $\mu $ denotes the magnetic moment of spinning particle. Using
relation Eq. (2), the total Lagrangian including the SO coupling Eq. (4) $%
\mathcal{L}%
=%
\mathcal{L}%
_{0}+%
\mathcal{L}%
_{i}$can be written as familiar
form,
\begin{equation}%
\mathcal{L}%
=%
\mathcal{L}%
_{0}+\frac{\mu }{c}[\mathbf{\dot{r}\cdot (S\times E)}]%
\end{equation}%
Noticing that the spin variable $\mathbf{S}$ is a function of $g$, we can
obtain equations of motion in the configuration space by means of the
variation  $\delta\int%
\mathcal{L}%
dt=0$ with respect to space coordinates $%
\mathbf{r}$ and spin-space coordinate $g$ respectively
\begin{eqnarray}
M\mathbf{\ddot{r}+\dot{p}}_{i}&  =&-\nabla V(r)+\frac{\mu }{c}\nabla \lbrack
\mathbf{S\cdot }(\mathbf{E\times \dot{r}})]  \label{6}\\
\mathbf{\dot{S}}&  =&\frac{\mu }{c}(\mathbf{\dot{r}\times E})
\times \mathbf{S}  \label{7}
\end{eqnarray}%
where
\[\mathbf{p}_{i}=\frac{\mu }{c}\mathbf{(S\times E)}\]
is called the intrinsic momentum\cite{Liang1}. In the absence of SO
coupling (i.e. $\mathbf{E=0}$), the spin variables become a conserved
quantity and the equation of motion Eq. (6) reduces to that of a point
particle in the central potential $V(r)$. The canonical momentum is defined
by
\begin{equation}
\mathbf{p=}\nabla _{\mathbf{\dot{r}}}
\mathcal{L}%
=M\mathbf{\dot{r}+}\frac{%
\mu }{c}\mathbf{(S\times E)}
\end{equation}%
Then the Hamiltonian has the usual minimum coupling form
\begin{equation}
H=\frac{[\mathbf{p-\frac{\mu }{c}A}]^{2}}{2M}\mathbf{+}V(r)  \label{9}
\end{equation}%
with an effective vector gauge-potential $\mathbf{A=S\times E}$, which, we
will see, becomes a pseudo non-Abelian gauge field after quantization. The
gauge field structure may be the most significant consequence of our
formalism for the introduction of classical spin-variables.
\subsection{Reduced classical equation of motion}

The coupled classical equations of motion (6) and (7) can be reduced (in the
initial condition $\dot{z}|_{t=0}=0$) to\cite{Liang2,Liang1}%
\begin{eqnarray}
M\ddot{x}& =-\frac{\partial }{\partial x}V(r)   \nonumber\\
M\ddot{y}& =-\frac{\partial }{\partial y}V(r)  \label{10}
\end{eqnarray}%
and
\begin{eqnarray}
\dot{S}_{x}&  =&-\frac{q}{r^{2}}[\dot{x}y-\dot{y}x]S_{y}  \nonumber \\
\dot{S}_{y}&  =&\frac{q}{r^{2}}[\dot{x}y-\dot{y}x]S_{x}  \nonumber \\
\dot{S}_{z}&  =&0  \label{11}
\end{eqnarray}%
up to the order $0$($q^{2}$), where $q=\eta \mu /c<<1$ is a dimensionless
small parameter, since the SO coupling is an effect of relativity. It is
interesting to see that the motion of central mass is effectively confined
in a 2D plane and is not affected by the spin motion. The solutions of
intrinsic (spin variables) motions are found as
\begin{eqnarray}
S_{x}(t)&  =&S_{x}(0)\cos [q\varphi (t)]+S_{y}(0)\sin [q\varphi (t)]  \nonumber \\
S_{y}(t)&  =&-S_{x}(0)\sin [q\varphi (t)]+S_{y}(0)\cos [q\varphi (t)]  \nonumber\\
S_{z}(t)&  =&S_{z}(0)  \label{12}
\end{eqnarray}
where $\varphi (t)=\arctan y(t)/x(t)$ and $S_{i}(0) (i=x,y,z)$ denotes the
initial value of spin variables. The motion of spin variables is nothing but
a Larmor precession around the $z$-axis with the $z$-component being
constant ( $S_{z}$=const). The precession angle $q\varphi (t)$ is
proportional to the polar angle of particle position $\varphi (t)$. For the
case of non-integer $q$ the spin vector does not return to its original
orientation even if the particle comes full circle. This , we will see, is
the dynamic reason of the non-Abelian anyon. The motion of central mass
along $z$-direction is a relativistic oscillating-perturbation around the $%
z=0$ plane with velocity%
\[\dot{z}=-\frac{q}{Mr}[S_{x}(0)\sin \varphi -S_{y}(0)\cos \varphi
]\]
which does not affect the planner motion of Eq. (\ref{10}) and the spin
precession of Eq. (\ref{11}) for periodic orbits in the central potential.
Thus we can consider only the motion of central mass in 2D space with
spin-orbit coupling.
\subsection{ Effective Hamiltonian in 2D spatial space and classical orbits}

\bigskip The effective Lagrangian corresponding to the reduced equations of
motion Eq. (10-11) in 2D spatial space of polar coordinate is seen to be $%
\mathcal{L}%
^{e}=%
\mathcal{L}%
_{0}^{e}+%
\mathcal{L}%
_{i}^{e}$ with $%
\mathcal{L}%
_{0}^{e}=\frac{M}{2}(\dot{r}^{2}+r^{2}\dot{\varphi }^{2})-V(r)$, where the SO coupling term is of the Wess-Zummino
type,

\begin{equation}
\mathcal{L}%
_{i}^{e}=qS_{z}\dot{\varphi}
\end{equation}
The canonical momentums conjugate to $r,\varphi$ are%
\begin{eqnarray}
p_{r} & =\frac{\partial%
\mathcal{L}%
^{e}}{\partial\dot{r}}=M\dot{r} \\
L_{z}^{c} & =L_{z}^{K}+qS_{z}%
\end{eqnarray}
$L_{z}^{c}$ is the canonical angular momentum (CAM), while $L_{z}^{K}$ $%
=Mr^{2}\dot{\varphi}$ is kinetic angular momentum (KAM). The Hamiltonian is%
\begin{equation}
H=\frac{p_{r}^{2}}{2M}+\frac{1}{2Mr^{2}}[L_{z}^{c}-qS_{z}]^{2}+V(r)
\label{16}
\end{equation}

Considering non-zero initial KAM of the value
\begin{equation}
L_{z}^{K}=\gamma \hbar
\end{equation}
the orbital motion of zero-energy satisfies the equation

\begin{equation}
\frac{(L_{z}^{K})^{2}}{2Mr^{4}}[(\frac{dr}{d\varphi })^{2}+r^{2}]-\frac{%
\varrho }{r^{2k+2}}=0  \label{18}
\end{equation}%
Changing variables successively to $\xi =r/a_{c},y=\xi ^{k}$ and $x=\cos
^{-1}y$, a direct mathematical solution\cite{Nieto} can be obtained by
integrating the orbital Eq. (\ref{18})
\begin{eqnarray}
\varphi -\varphi _{0}& =\int_{1}^{\xi }\frac{\xi ^{k-1}}{\sqrt{1-\xi ^{2k}}}%
d\xi =k^{-1}\int_{1}^{y}\frac{dy}{\sqrt{1-y^{2}}} \nonumber\\
& =-k^{-1}\int_{0}^{\arccos y}dx\nonumber
\end{eqnarray}%
and the result is%
\[\xi ^{k}=\cos k(\varphi -\varphi _{0})\]
or
\begin{equation}
r^{k}=a_{c}^{k}\cos [k(\varphi -\varphi _{0})]  \label{19}
\end{equation}%
where $a_{c}^{2k}={2M\varrho/(L_{z}^{K})^{2}}={2M\varrho}/
{\gamma ^{2}\hbar ^{2}}$. The classical orbits with the initial angle setting
to zero ($\varphi _{0}=0$ ) are shown in Fig. 1 for the closed orbits with
the scalar potentials of power-indexes $k=1,7/3,4,9/2,5$, $17/3$, and in
Fig. 2 for the open orbits with potential power-indexes $k=$ $-6$ ,$-17/4$,
where we set the dimensionless constant $2M\varrho /\hbar ^{2}=1$ in the
numerical calculation. The classical orbits Eq. (\ref{19}) are invariant
under a rotation-angle $2\pi /\left\vert k\right\vert $ and the $2\pi $%
-symmetry holds only for $k=1$. The initial values of KAM for closed orbits
in Fig. 1 are set by $\gamma =15,35,60,135/2,85,75,$ respectively, while $%
\gamma $= $255/4$, $90$ in Fig. 2.
\section{\protect\bigskip Quantum wave functions and QCC}

\subsection{Pseudo non-Abelian gauge field}

After quantization the gauge field in Hamiltonian Eq. (\ref{9}) becomes a
field-valued operator $\mathbf{A=S\times E}$, with%
\begin{equation}
\mathbf{S=\hbar }\frac{\mbox{\boldmath$\sigma$}}{2}
\end{equation}%
for the spin-1/2 case, which is a matrix field called \ pseudo non-Abelian
gauge field. The\textbf{\ }Hamilton operator in the representation of a
covariant derivative can be written as%
\begin{equation}
H=-\frac{\hbar ^{2}}{2M}\sum_{i=x,y,z}D_{i}^{2}\mathbf{+}V(r)
\end{equation}%
where
\begin{equation}
D_{i}=\frac{\partial }{\partial x_{i}}-i\beta A_{i}
\end{equation}%
denotes a covariant derivative with $\beta =\frac{\mu }{2c}$. The
non-Abelian gauge field%
\begin{equation}
A_{i}=\sum_{j,k=x,y,z}\epsilon _{ijk}\sigma _{j}E_{k}
\end{equation}%
(where $\epsilon _{ijk}$ is the Levi-Civita tensor) has the explicit form%
\begin{eqnarray}
A_{x}&  =&-\frac{\eta \sin \varphi }{r}\sigma _{z}  \nonumber \\
A_{y}&  =&\frac{\eta \cos \varphi }{r}\sigma _{z}  \nonumber \\
A_{z}&  =&\frac{\eta }{r}(\sin \varphi \sigma _{x}+\cos \varphi \sigma _{y})
\end{eqnarray}%
for the cylindrically symmetric electric field Eq. (\ref{3}). The effective
magnetic-field can be expressed by the antisymmetric tensor%
\begin{equation}
F_{i,j}=\frac{i}{\beta }[D_{i},D_{j}]=(\frac{\partial A_{j}}{\partial x_{i}}-%
\frac{\partial A_{i}}{\partial x_{j}})-i\beta \lbrack A_{i},A_{j}]
\end{equation}%
which gives rise to
\begin{eqnarray}
B_{x}&  =&0  \nonumber \\
B_{y}&  =&\frac{\eta }{r^{2}}\{(1-\beta \eta )\sin 2\varphi \sigma
_{x}+[(1-\beta \eta )\cos 2\varphi +\beta \eta ]\sigma _{y}\}  \nonumber \\
B_{z}&  =&\frac{\eta }{r^{2}}\{[(1-\beta \eta )\cos 2\varphi -\beta \eta
]\sigma _{x}-(1-\beta \eta )\sin 2\varphi \sigma _{y}\}
\end{eqnarray}%
The unitary operator of gauge transformation is defined as%
\begin{equation}
U=e^{i\beta \sum_{i}f_{i}\sigma _{i}}
\end{equation}%
where $f$ is a function of space coordinate. The covariant derivative
operator under the gauge transformation becomes%

\[D_{i}^{^{\prime }}=UD_{i}U^{\dag }=\frac{\partial }{\partial x_{i}}-i\beta
A_{i}^{^{\prime }}\]

where gauge field becomes%

\[A_{i}^{^{\prime }}=UA_{i}U^{\dag }-\frac{i}{\beta }(\frac{\partial U}{%
\partial x_{i}})U^{\dag }\]

Spinor wave function transformation is
\begin{equation}
\psi ^{^{\prime }}=U\psi
\end{equation}%
It is worth while to remark that the non-Abelian gauge field in quantum
mechanics is only of formal meaning which is a concept first introduced by
Wilczek and Zee\cite{Wilczek}. To avoid confusion we call it the pseudo
non-Abelian gauge field. The SO coupling and non-Abelian gauge field have
become an active research field of ultracold atoms in optical fields. The
gauge field method might help us to have a deeper understanding of the
fundamental quantum phenomena in the atomic, molecular and optical physics.
\subsection{2D wave functions, fractional angular momentum and QCC}

We are particularly interested in the QCC. To this end we turn to the
quantum mechanics of reduced 2D problem with Hamiltonian Eq. (\ref{16}). In
the polar coordinate, the stationary Schr\"{o}dinger equation of zero-energy
with wave function $\Psi =R(r)Y_{j,\chi }(\varphi )$

\[\frac{-\hbar ^{2}}{2M}\{\frac{\partial ^{2}}{\partial r^{2}}+\frac{1}{r}%
\frac{\partial }{\partial r}+\frac{1}{r^{2}}[\frac{\partial }{\partial
\varphi }-i\frac{1}{2}q\hat{\sigma}_{z}]^{2}\}\Psi +V(r)\Psi =0  \]

becomes%
\begin{eqnarray}
(\frac{\partial ^{2}}{\partial r^{2}}+\frac{1}{r}\frac{\partial }{\partial r}%
-\frac{j^{2}}{r^{2}})R_{j}(r)+\frac{2M}{\hbar ^{2}}\frac{\varrho }{r^{2k+2}}%
R_{j}(r)&  =&0 \\
(\frac{\partial }{\partial \varphi }-i\frac{1}{2}q\hat{\sigma}%
_{z})^{2}Y_{j,\chi }(\varphi )&  =&-j^{2}Y_{j,\chi }(\varphi )
\end{eqnarray}%
where $\chi $ denotes the spin index and $j$ is the eigenvalue of orbital
angular-momentum. The angular wave function, which possesses a non-Abelian
topological phase, is seen to be
\begin{equation}
Y_{j,\chi }(\varphi )=Y_{j}(\varphi )e^{i(q\hat{\sigma}_{z}/2)\varphi }|\chi
\rangle
\end{equation}%
with $Y_{j}(\varphi )=C_{j}e^{ij\varphi }$ where $C_{j}=\sqrt{\left\vert
k\right\vert /2\pi }$ is the normalization constant depending on the angular
momentum quantization in 2D spatial space and $|\chi \rangle $ denotes a
spin state. We consider the QCC with the meaning that both the probability
density of wave functions and the expectation value of spin-operator well
coincide with the classical orbits Eq. (\ref{19}) and the classical spin
precession Eq. (\ref{12}) respectively. This kind of QCC results in a
special boundary condition of the angular eigenstates such that the
rotational period of angular wave-functions, which is not necessarily 2$\pi $%
, should be the same as that of classical orbits i.e. $Y_{j}(\varphi
)=Y_{j}(\varphi +\frac{2\pi }{\left\vert k\right\vert })$. Thus the CAM
eigenvalue is no longer integer but should be set as
\begin{equation}
j=\nu \left\vert k\right\vert
\end{equation}%
where $\nu $ is a non-vanishing integer. Introducing the dimensionless
radius $z={\tilde{a}_{q}}/{r}$ and a parameter $\tilde{a}_{q}$ with
dimension of length the radius equation,
\[\
(\frac{\partial ^{2}}{\partial r^{2}}+\frac{1}{r}\frac{\partial }{\partial r}%
-\frac{j^{2}}{r^{2}})R_{j}(r)+\frac{2M}{\hbar ^{2}}\frac{\varrho }{r^{2k+2}}%
R_{j}(r)=0
\]
becomes%
\[
\lbrack z^{2}\frac{d^{2}}{dz^{2}}+z\frac{d}{dz}+B^{2}z^{2k}-j^{2}]\Theta
_{j}(z)=0
\]
where $B^{2}={2M\varrho }/{\hbar ^{2}\tilde{a}_{q}^{2k}}$, and can be
rewritten as%
\begin{equation}
\{y^{2}\frac{d^{2}}{dy^{2}}+y\frac{d}{dy}+[y^{2}-\nu ^{2}]\}\Theta _{j}(y)=0
\end{equation}%
by the change of variables $\xi =z^{k}$ and $y={B}\xi/{k} $. The
solution is Bessel function of order $\nu $,
\[\
\Theta _{j}(y)=N_{j}J_{\nu }(y)
\]
which is the same as in Ref. \cite{Makowski2}. The radius wave function is%
\[
R_{j}(r)=N_{j_{{}}}J_{\nu }(\frac{1}{\left\vert k\right\vert r^{k}})
\]
With the angular quantum-number $j=\nu \left\vert k\right\vert $, classical
orbits can be mimicked by the probability density of wave functions, which
we will see in the followings. The normalization constant obtained from the
integral
\[\
\int_{0}^{\infty }R_{j}^{2}(r)rdr=1
\]
with the infinite integrals of Bessel function\cite{Watson} is%
\begin{equation}
\left\vert N_{j}\right\vert ^{2}=2\sqrt{\pi }\left\vert {k}\right\vert ^{%
\frac{2}{k}+1}\frac{\Gamma (\frac{1}{k}+1)\Gamma (\frac{1}{k}+\nu +1)}{%
\Gamma (\frac{1}{k}+\frac{1}{2})\Gamma (\nu -\frac{1}{k})}
\end{equation}%
by setting $B\tilde{a}_{q}^{k}=1$. It is of fundamental importance to show
the correspondence between wave function and classical orbit. Recently, it
was demonstrated that the probability density of wave function can be well
localized on the classical orbits in a 2D central potentials with the
representation of $SU(2)$ coherent states\cite{Makowski2,Makowski3}, which%
\textbf{\ }is a superposition of degenerate eigenstates. To coincide with
the classical orbits\textbf{\ }we construct the macroscopic wave functions
in the spirit of $SU(2)$ coherent-state representation in terms of the
degenerate partial-waves and the macroscopic wave-function is
\begin{equation}
\Psi _{n}=\frac{1}{2^{n/2}}\sum_{\nu =0}^{n}{n\choose \nu}^{1/2}R_{j}(r)Y_{j,\chi }(\varphi )  \label{35}
\end{equation}%
which is normalized to unity in the angular range of $\varphi $ being from $%
-\pi /\left\vert k\right\vert $ to $\pi /\left\vert k\right\vert $. The
probability densities\ $\Psi _{n}^{\dag }\Psi _{n}$ of macroscopic state
with $n=30$ shown in Figs. 3 and 4 are in excellent agreement with the
classical orbits (see Figs.1,2) indicating the exact QCC\textbf{. }As a
consequence we have the fractional orbital-angular-momentum with the
eigenvalue $j=\nu \left\vert k\right\vert $, if $\nu \left\vert k\right\vert
$ is a fractional number.

Besides the spatial orbits the QCC holds also for the spin motion. To this
end we begin with the expectation value of Heisenberg equation of the spin
operator in the macroscopic wave function Eq. (\ref{35}), which gives rise to%
\begin{eqnarray*}
\langle \dot{\sigma }_{x}\rangle &  =&q\dot{\varphi
\langle }\sigma _{y}\rangle \\
\langle \dot{\sigma }_{y}\rangle &  =&-q\dot{\varphi
\langle }\sigma _{x}\rangle \\
\langle \dot{\sigma }_{z}\rangle &  =&0
\end{eqnarray*}%
where%
\[\
\langle \sigma _{i}\rangle =\langle \chi |\sigma _{i}|\chi \rangle
\]
and the constant being angular velocity is calculated as
\[\
\dot{\varphi }=\langle \psi _{n}|\frac{\hat{L}_{z}^{K}}{Mr^{2}}%
|\psi _{n}\rangle
\]
in which $\psi _{n}=\frac{1}{2^{n/2}}\sum_{\nu =0}^{n}{n\choose \nu}^{1/2}R_{j}(r)Y_{j}(\varphi )$ is the macroscopic orbital-wave-function and $%
\hat{L}_{z}^{K}$ is the orbital angular momentum operator. The expectation
value of spin equation of motion in the macroscopic state is the same as the
classical one Eq. (\ref{11}).
\subsection{Non-Abelian Anyon}

\bigskip The orbital angular momentum operator is\textbf{\ }$\hat{L}%
_{z}^{K}=-i\hbar \partial /\partial \varphi $ and total angular-momentum
operator i.e the CAM operator reads%
\begin{equation}
\hat{L}_{z}^{c}=-i\hbar \frac{\partial }{\partial \varphi }+\frac{1}{2}%
q\hbar \hat{\sigma}_{z}
\end{equation}%
which possesses eigenstates $Y_{j,\chi }(\varphi )$ if the spin state is
chosen as eigenstates of $\hat{\sigma}_{z}$, such that $|\chi \rangle =|\pm
\rangle $, where $\hat{\sigma}_{z}|\pm \rangle =\pm |\pm \rangle $. Thus the
total angular-momentum spectrum $\Lambda $ is seen to be%
\begin{equation}
\Lambda =(\nu \left\vert k\right\vert \pm \frac{1}{2}q)\hbar
\end{equation}%
with $\nu $ being a integer. The angular momentum is quantized with an
eigenvalue-space $\left\vert k\right\vert $, which can be a fractional
number determined completely by the QCC. The gauge potential of SO coupling
can shift the spectrum of CAM by an arbitrary value $q\hbar/{2} $,
which results in a common topological phase for all wave functions in the
given model, since $q$ is not necessarily an integer. Thus non-Abelian
anyons\cite{Wliczek1} naturally appear, which obey non-Abelian braiding
statistics.

\section{Conclusion}

In summary, both classical and quantum periodic orbits of a SO coupling
model in a wide class of central potentials are obtained analytically, where
the classical spin-variables are introduced in a consistent way to give rise
to the correct equation of motion. The exact QCC with a perfect match
between classical orbits and the spatial probability clouds is found with
the coherent superposition of angular wave functions, which imposes a
special boundary condition on the angular wave functions resulting in the
fractional angular momentum quantization. Simultaneously, the QCC for the
spin degree of freedom is also established, which displays the exactly same
spin-precession in addition to the orbital motion of central mass. The SO
coupling model exhibits explicitly a pseudo non-Abelian gauge field and the
anyon behavior.

\section*{Acknowledgment}

This work was supported by National Nature Science Foundation of China
(Grant No.11075099).

\newpage

\begin{figure}
\centering \vspace{0mm} \hspace{0cm} \scalebox{0.65}
{\includegraphics{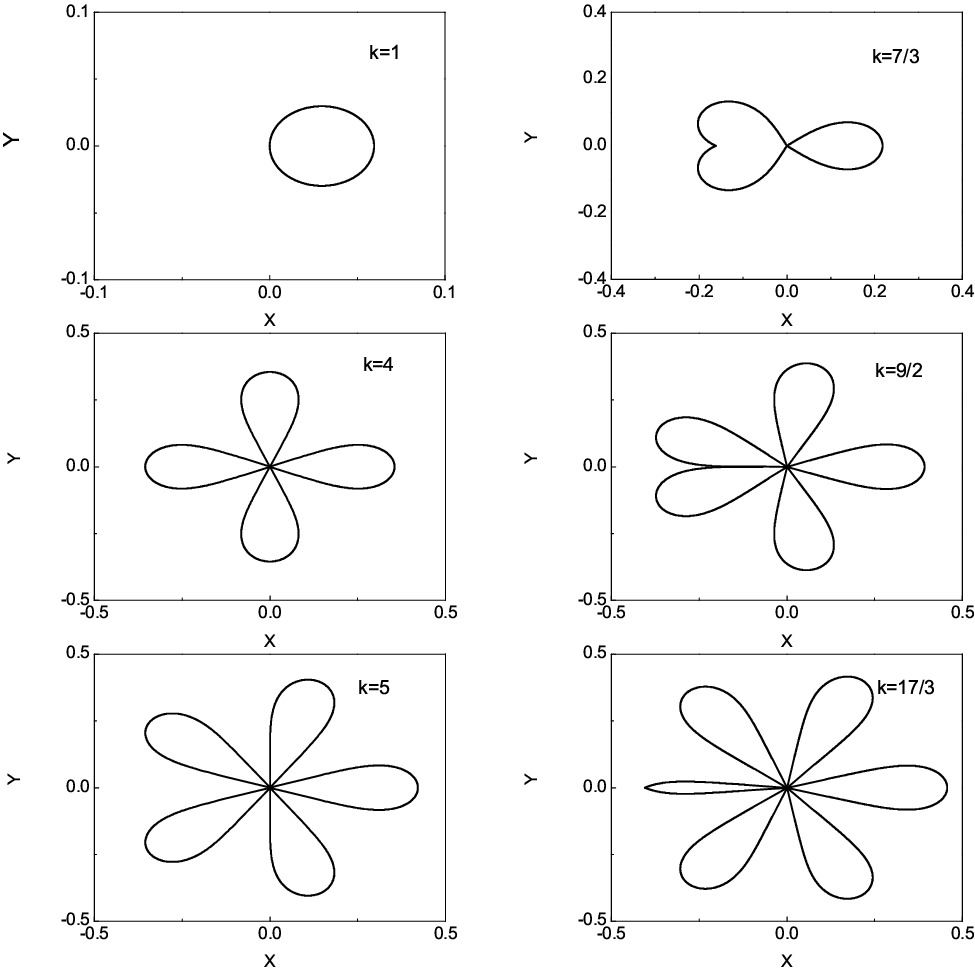}} \caption{Closed
classical periodic-orbits for various value of KAM with potential
power-indexes $k=1,7/3,4,9/2,17/3,5$.}
\end{figure}

\begin{figure}
\centering \vspace{0mm} \hspace{0cm} \scalebox{0.65}
{\includegraphics{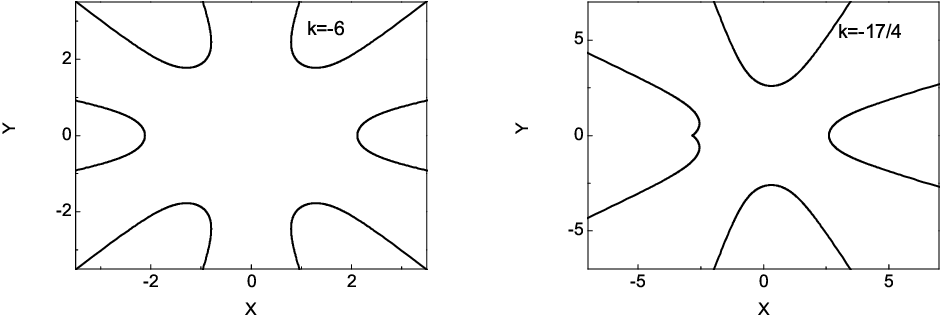}} \caption{Open classical periodic
orbits for various value of KAM with the potential power-indexes $k=-6,-17/4$.}
\end{figure}

\begin{figure}
\centering \vspace{0mm} \hspace{0cm} \scalebox{0.8}
{\includegraphics{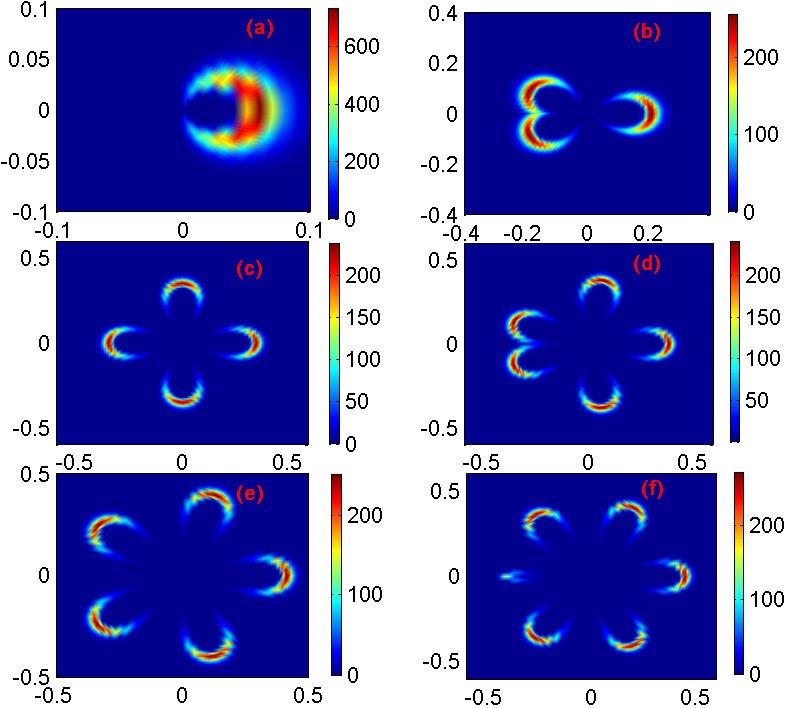}} \caption{(color online) Probability
density $\left\vert Y_{j,\chi}(\varphi)\right\vert ^{2}$ for the parameters
corresponding to closed classical orbits shown in Fig. 1.}
\end{figure}

\begin{figure}
\centering \vspace{0mm} \hspace{0cm} \scalebox{0.8}
{\includegraphics{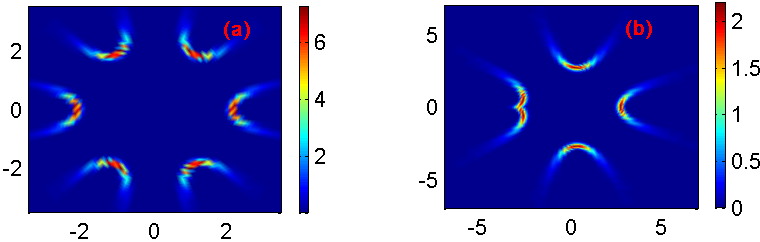}} \caption{(color online) Probability
density $\left\vert Y_{j,\chi}(\varphi)\right\vert ^{2}$ for the parameters
corresponding to open classical orbits shown in Fig. 2.}
\end{figure}

\end{document}